\documentclass[twocolumn,showpacs,superscriptaddress]{revtex4-1}  
\usepackage{graphicx}  

\usepackage{dcolumn}   
\usepackage{bm}        
\usepackage{amssymb}   
\usepackage{amsmath}

\begin{document}


\title{Generation of  an optical Schr{\"o}dinger-cat-like state in a non-ideal cavity by injecting opposite-phase atomic dipoles}

\author{Daeho Yang}
\affiliation{Department of Physics and Astronomy, Seoul National University, Seoul 08826, Korea} 
\author{Junki Kim}
\affiliation{Department of Physics and Astronomy, Seoul National University, Seoul 08826, Korea} 
\author{Moonjoo Lee}
\affiliation{Institut f\"{u}r Experimentalphysik, Universtit\"{a}t Innsbruck, Technikerstra$\beta$e 25, 6020 Innsbruck, Austria}
\author{Young-Tak Chough}
\affiliation{Department of Medical Technology, Gwangju University, Gwangju 61743, Korea}
\author{Kyungwon An}
\email{kwan@phya.snu.ac.kr}
\affiliation{Department of Physics and Astronomy, Seoul National University, Seoul 08826, Korea} 
\date{\today}

\begin{abstract}
We propose a method to generate  an optical Schr\"{o}dinger-cat-like state in a cavity in a substantial decoherence regime.
Even when the cavity decay rate is considerably large, a cat-like state can be generated in a laser-like setting if the gain for the field is larger than the loss.
Under the condition that opposite-phase atomic dipoles repeatedly traverse the cavity, the cavity field converges to a squeezed vacuum state in a steady state.
A Schr\"{o}dinger-cat-like state is then generated when a single photon decay occurs.
The phase-space distribution of the cat state can be revealed in homodyne detection by using the decaying photon as a herald event.
Quantum trajectory simulation was used to identify the conditions for the Schr\"{o}dinger-cat-like state formation as well as to analyze the properties of those states.
Based on these simulations, possible experiments are proposed within the reach of the current technology.
\end{abstract}


\maketitle

\section{Introduction}
Quantum mechanics allows simultaneous existence of two or more states at once as a superposition state. 
Its physical meaning was one of the key issues in early days of quantum mechanics.
Especially, the Schr{\"o}dinger's cat -- a superposition state of macroscopic states -- has drawn much attention due to its counterintuitive aspects.
Diverse theoretical and experimental studies were performed to realize Schr{\"o}dinger-cat-like states in various physical systems.
As a result, superposition states of atomic external and internal states \cite{Atomic cat-1,Atomic cat-2, ion cat_new}, microwave photons \cite{Brune-92, Del-08, Microwave-circuit, deterministic_cat} and optical photons \cite{Kitten, Neergaard-Nielsen-06, Takahashi-08, Ourjoumtsev-07} have been realized experimentally.

In an optical system, a large coherent state can be regarded as a classical state due to its well-defined phase and intensity.
In this manner, a superposition of coherent states (SpCS) is often referred to as a Schr{\"o}dinger's cat state: one coherent state is a live cat and the other coherent state with an opposite phase is a dead cat.
SpCSs are known to be useful in quantum information processing \cite{QI-1,QI-2,QI-3} as well as in quantum metrology \cite{Quantum metrology-1,Quantum metrology-2,Quantum metrology-3}.
SpCSs investigated so far can be classified as free-propagating  or bounded SpCS.
Generation of free-propagating SpCSs has been analyzed
by subtracting photon(s) 
from a squeezed vacuum at once \cite{Kitten, Dakna-97,Neergaard-Nielsen-06} as well as in a time-separated way \cite{Takahashi-08,time-separated theory}, adding photons to a squeezed vacuum \cite{Marek-08} and performing homodyne detection on a photon number state \cite{Ourjoumtsev-07}.
Bounded SpCSs have been generated by inducing a phase shift on a cavity field by dispersive atom-field interaction \cite{Brune-92, Del-08} and by utilizing conditional qubit rotation as well as conditional cavity displacement \cite{deterministic_cat}.

It is noteworthy that generation of SpCSs in cavities has been achieved only in ultralow-loss systems such as superconducting cavities.
It is because a SpCS in a cavity experience decoherence at a rate proportional to $\sim \kappa \langle n \rangle$, where $\kappa$ is the cavity decay rate and $\langle n\rangle$ is the mean photon number.
Moreover, in order to realize a SpCS in a cavity the previous studies utilized well-controlled atom/qubit operations, which should be completed well before the system undergoes significant decoherence. 
Therefore, ultralow cavity loss was essential in those previous studies.

In this paper, we propose a method to generate an optical cat-like state in a cavity even in the presence of substantial cavity decay.
Our proposal is based on the cavity-QED microlaser system \cite{microlaser-1,microlaser-2,microlaser-3, nanohole-14} with opposite-phase atomic dipoles traversing the cavity repeatedly.
Here atomic dipoles are referred to two-level atoms in a superposition state of the ground and excited states.
Under the condition that the gain for the field is larger than the loss, the cavity field converges to a squeezed vacuum state (SqVS) due to the interference caused by the opposite-phase atomic dipoles.
Once a SqVS is formed in a steady state, it collapses to a Schr\"{o}dinger-cat-like state when a single photon decay occurs. Note that the cavity decay is even essential in this scheme.
Further interaction between the atoms and the field then restores the SqVS in the cavity again in a characteristic time, inversely proportional to the gain, and the whole process can be repeated.
Because the formation of the SqVS depends mainly on the ratio of gain and loss, a Schr\"{o}dinger-cat-like state can be generated even under substantial cavity decay as long as the gain exceeds the loss just like a laser. 
Our scheme is advantageous in speed compared to the other previous methods of generating bounded cat-like states, and therefore it can be useful when a high repetition rate is preferred.
In contrast to free-propagating SpCSs of broad bandwidths, out scheme provides a narrow bandwidth, which would enable interactions between cat-like states and two-level systems. Moreover, as an intracavity cat-generation scheme, it also allows systematic investigation of decoherence as well as quantum feedback for cat-state stabilization.

This paper is organized as follows.
In Sec.\ \ref{sec2}, we briefly introduce our physical system and explain how a SqVS is formed. 
In Sec.\ \ref{sec3}, we assume an infinitesimally small cavity decay and then show a photon decay induces a quantum jump from a SqVS to a cat-like state. 
In Sec.\ \ref{sec4}, we consider a case with substantial cavity decay by using quantum trajectory simulation (QTS). 
We then show in Sec.\ \ref{sec5} that the SqVS is restored in a certain characteristic time and the process can be repeated. In Sec.\ \ref{sec6}, we suggest possible experiments with parameters confirmed by QTS to be in the reach of the current technology.
We then summarize our study and draw a conclusion in Sec.\ \ref{sec7}.
Analytic approaches were employed in Secs.\ \ref{sec2} and \ref{sec3} to find the characteristic time and to show the formation of a cat-like state. Numerical approaches were used in Secs.\ \ref{sec3}, \ref{sec4}, \ref{sec5} and \ref{sec6} to analyze the effect of cavity decay as well as the properties of the formed cat-like state.

\section{Generating a squeezed vacuum state in a lossless cavity with opposite-phase atomic dipoles}\label{sec2}

For simplicity, we neglect the cavity decay in this section. We also assume that two atoms prepared in superposition states traverse a cavity simultaneously 
for now although this simultaneous traverse condition is not mandatory as to be discussed below. We assume that
one atom is prepared in a superposition state of the ground and excited states with a certain phase.
The other atom is prepared in another superposition state with the same ground and excited state probability but with an opposite phase to that of the first atom. The state of these atoms can be written as
\begin{equation}
(\alpha \left| \downarrow \right\rangle + \beta \left| \uparrow \right\rangle)(\alpha \left| \downarrow \right\rangle - \beta \left| \uparrow \right\rangle) .
\end{equation}
where $\left| \downarrow \right\rangle$ and $\left| \uparrow \right\rangle$ are the ground and excited states of atom, respectively, and $\left|\alpha\right|^2$ and $\left|\beta\right|^2$ are the probability of atom in the ground and excited states, respectively.
The atoms interact with the cavity field with the same atom-field coupling constant $g$ for a time duration of $t_{\rm int}$.
As soon as they leave the cavity, a pair of newly prepared atoms enter the cavity and interact with the cavity field for another $t_{\rm int}$.
This process is repeated until a stationary field state is reached.

In this case, the atom-field interaction is described by the following Tavis-Cummings Hamiltonian.
\begin{equation}
H_{\rm int}=-ig\left[a(\sigma_1^\dagger+\sigma_2^\dagger)+a^\dagger(\sigma_1+\sigma_2)\right] ,
\end{equation}
where $a(a^\dagger)$ is the photon annihilation(creation) operator and $\sigma_i$($\sigma_i^{\dagger}$) is the lowering(raising) operator of the $i$th atom. 
The cavity frequency is assumed to be on resonance with the atomic transition. 
We also assume that the coupling constant is much larger than the atomic spontaneous emission rate, so the atomic spontaneous emission to the environment is neglected in our analysis throughout this paper.
By using a shorthand notation $ \left|\uparrow\uparrow\right\rangle \equiv \left|\uparrow\right\rangle|\left\uparrow\right\rangle$, $ \left|\uparrow\downarrow\right\rangle \equiv \left|\uparrow\right\rangle|\left\downarrow\right\rangle$, etc, the state of two opposite-phase atoms can be expressed as
\begin{equation}
\alpha^2 \left| \downarrow\downarrow \right\rangle - \beta^2 \left| \uparrow \uparrow \right\rangle + \alpha\beta \left[ \left| \uparrow\downarrow \right\rangle - \left| \downarrow\uparrow \right\rangle \right] .
\end{equation}
Under the Tavis-Cummings Hamiltonian, the term enclosed with the square brackets, a singlet (subradiant) state, does not interact with the cavity field. 
Only the first two terms interact with the cavity field. 

In order to find the steady state under the condition that atom pairs repeatedly traverse the cavity, let us consider the change of the cavity field with time under the Tavis-Cummings Hamiltonian. By writing the atom-field combined state as $\left|\psi\right\rangle$ and the field state as $\Sigma_{n=0}^{\infty} C_n|n\rangle$, the rate of change of the atom-field combined state at the moment of atomic injection can be described as 
\begin{eqnarray}
\frac{d|\psi\rangle}{dt} = -i &g& \left[ a( \sigma_1^\dagger+\sigma_2^\dagger)+a^\dagger(\sigma_1+\sigma_2) \right] \nonumber \\
& &\left[\alpha^2 \left| \downarrow\downarrow \right\rangle - \beta^2 \left| \uparrow \uparrow \right\rangle \right] \sum\limits_{n=0}^{\infty} C_n \left|n\right\rangle,
\end{eqnarray}
which is simplified to
\begin{equation}
\frac{d|\psi\rangle}{dt}=-ig(\left|\uparrow\downarrow\right\rangle+\left|\downarrow\uparrow\right\rangle)\left[(\alpha^2a-\beta^2a^\dagger)\sum\limits_{n=0}^{\infty} C_n\left|n\right\rangle\right].
\label{steady}
\end{equation}
For the field state satisfying [ $\cdots$ ] $=0$, the cavity field remains unchanged afterwards. Interestingly, the state satisfying [ $\cdots$ ] $=0$ is an ideal SqVS with a squeezing parameter $r=\tanh^{-1} (\beta^2/\alpha^2)$. Note we obtain a stationary cavity field state even though we do not include the cavity decay here. The formation of a stationary state without cavity decay is due to the equilibrium between the excited and ground states of atoms: the excited state emits a photon and the ground state absorbs a photon simultaneously so there is no net change. Explicit derivation of the steady state formation under the assumption $gt_{\rm int} \ll 1$ is presented in Appendix \ref{appen:A}.

If we consider the interacting parts only, the initial atomic state can be understood as a two-atom squeezed state \cite{Squeezed_spin}.
A multi-atom (spin) squeezed state in a spinor notation can be understood as a state having reduced spin noise in one direction at the cost of increased spin noise in the other direction.
The squeezed state transfer from an atomic squeezed state to a field squeezed state was studied before \cite{Squeezed_spin, Squeezed atom-1,Squeezed atom-2}.
Most of the studies focused on the trapped atom cases, yielding a time-dependent squeezing parameter and requiring a large atomic squeezed state in order to achieve a substantial field squeezed state.
In our case, however, repeatedly injecting a two-atom squeezed state with a low atomic squeezing parameter
yields a large stationary squeezing parameter for the field.
Interestingly, while the usual optical parametric oscillation often used to generate intracavity field squeezing is limited to 3dB quadrature squeezing \cite{squeezing_limit_cavity}, there is no such limitation in our scheme based on opposite-phase atomic dipoles.

In this section, we have assumed that two opposite-phase atoms are traversing the cavity at once. As to be shown in Sec.~\ref{sec3}, \ref{sec4}, \ref{sec5} and \ref{sec6}, however, simultaneous traverse condition is not mandatory in fact.
Generalization to a non-simultaneous traverse case is 
analytically done in Appendix \ref{appen:B}.

\section{Inclusion of infinitesimally small cavity decay} \label{sec3}

\begin{figure}
\includegraphics[width=3in]{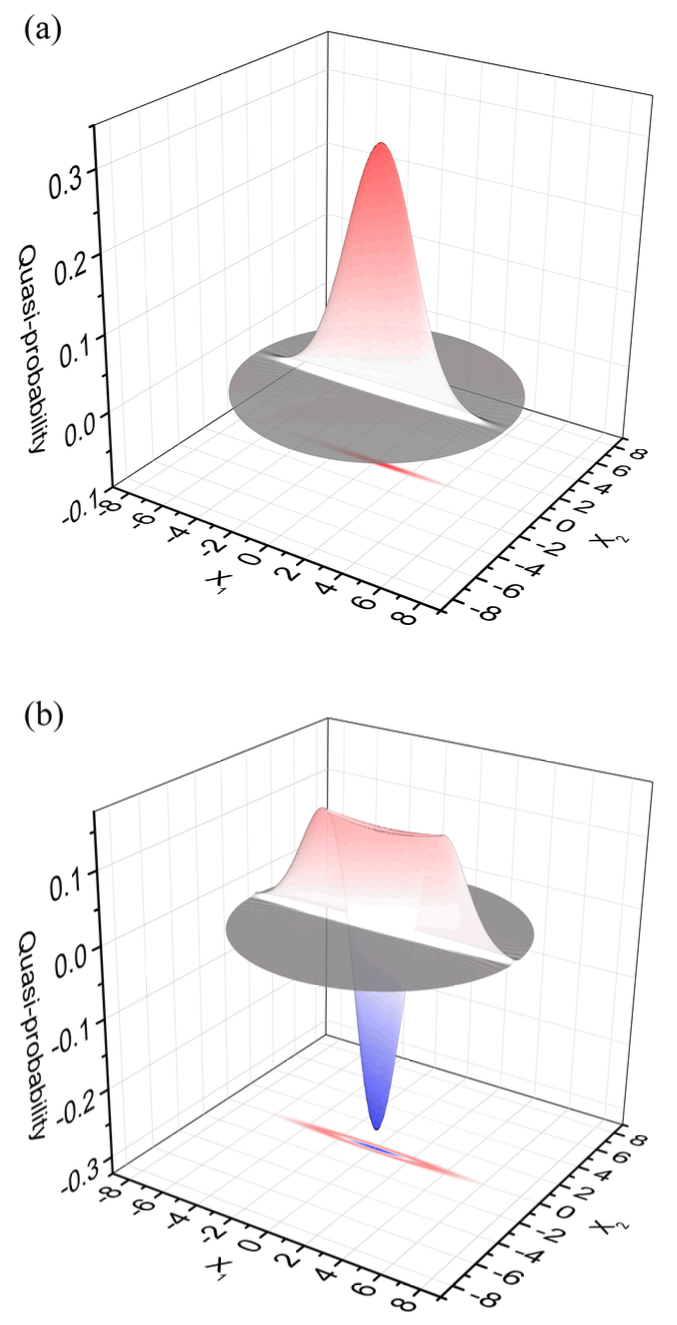}
\caption{(Color online) (a) Wigner distribution of the cavity field before a photon click. The SqVS shows 5.7dB squeezing in $X_2$ direction. Here, $X_1$ and $X_2$ are the quadratures, $(a+a^\dagger)/2$ and $(a-a^\dagger)/2i$, respectively. (b) Wigner distribution of the cavity field after a photon click. The SPSb-SqVS shows negative quasi-probability, a signature of a non-classical state. The state also shows 3.3dB squeezing in $X_2$ direction. The average photon number $\left\langle n \right\rangle$ is 3.0 in (a) and 10.0 in (b).  
The plots are  obtained from QTS results with the simulation conditions $\beta^2=0.46$ and $\kappa\simeq0$. Detailed information on our QTS is given in Sec.\ \ref{sec4}.}
\label{fig1}
\end{figure}

In this section, we consider the collapse of the cavity field due to a photon decay. 
To simplify the argument, we assume that the cavity decay rate is small enough not to disturb the formation of the SqVS.
Because of the cavity decay, a photon leaks out of the cavity and a photon click occurs on a photodetector outside.
Upon a photon click, a photon must then be subtracted from the cavity field and the wave function for the cavity field collapses.
A single-photon-subtracted state from a SqVS $|\xi\rangle$, or a single-photon-subtracted squeezed vacuum state (SPSb-SqVS) in short, in the cavity can be described by \cite{Biswas-07}
\begin{equation}
|\xi\rangle \longrightarrow a|\xi\rangle.
\end{equation}
It has been known that the state obtained by subtracting a single photon (or multiple photons) from a SqVS is a Schr\"{o}dinger-cat-like state in the studies of free-propagating SpCSs \cite{Dakna-97, Marek-08, Takahashi-08}.
Likewise, we observe a cat-like state in the cavity when a photon click occurs.
The Wigner distribution of  the cavity field before and after a photon click is shown in Fig.\ \ref{fig1}.

\begin{figure}
\includegraphics[width=3.4in]{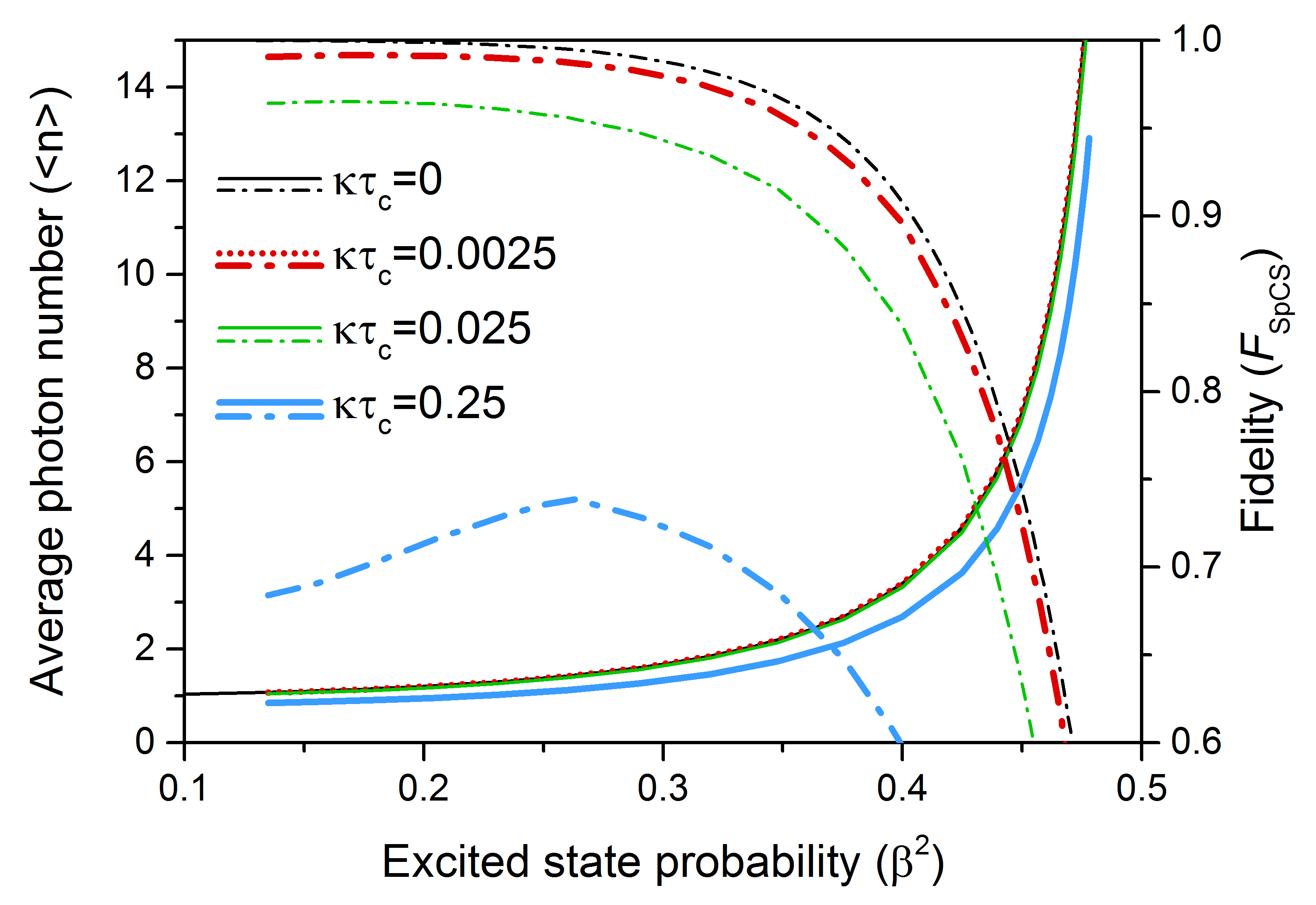}
\caption{(Color online) SPSb-SqVS's average photon number $\langle n \rangle$ (solid lines or dotted lines) and fidelity $F_{\rm SpCS}$ (dash-dot lines) with respect to a SpCS as a function of excited state probability $\beta^2$ of atoms.
As denoted in the legend, different colors (grayscales) and line styles represent different cavity decay rate $\kappa$ times the characteristic time $\tau_c$, which is defined in Eq.\ (6) in the text. 
For a small cavity decay ($\kappa\tau_c\ll 0.01$), $\langle n \rangle$ and $F_{\rm SpCS}$ only depends on the pumping parameter $\beta^2$. As the cavity decay rate increases, $\langle n \rangle$ decreases under the same pumping parameter. Plots are obtained from the QTS results under the condition $gt_{\rm int}=0.1$.}
\label{fig2}
\end{figure}

As studied in a previous paper \cite{Marek-08}, a SPSb-SqVS is more similar to a squeezed superposition of coherent states (Sq-SpCS) than a SpCS.
As a result, the fidelity $F_{\rm SpCS}=\left|\left \langle \psi |\psi_{\rm SpCS} \right\rangle\right|^{2}$ between a SPSb-SqVS of wavefunction $|\psi\rangle$ and a SpCS of wavefunction $|\psi_{\rm SpCS} \rangle$  is less than unity.
Moreover, 
the present SPSb-SqVSs have maximum fidelities with respect to odd-cat states. Every fidelity shown below is computed with respect to odd SpCS or odd Sq-SpCS.
Figure \ref{fig2} shows the average photon number of the SPSb-SqVS and its fidelity $F_{\rm SpCS}$ as a function of the excited state probability $\beta^2$ of atoms under the condition of an infinitesimally small cavity decay ($\kappa\simeq0$ with $\kappa$ the cavity decay rate).
For fair comparison, the fidelity $F_{\rm SpCS}$ is calculated between the SPSb-SqVS and the SpCS having the same average photon number as the SPSb-SqVS. 
The fidelity decreases as the excited state probability increases due to the increase in the squeezing parameter of the SPSb-SqVS state. 

\begin{figure}
\includegraphics[width=3.4in]{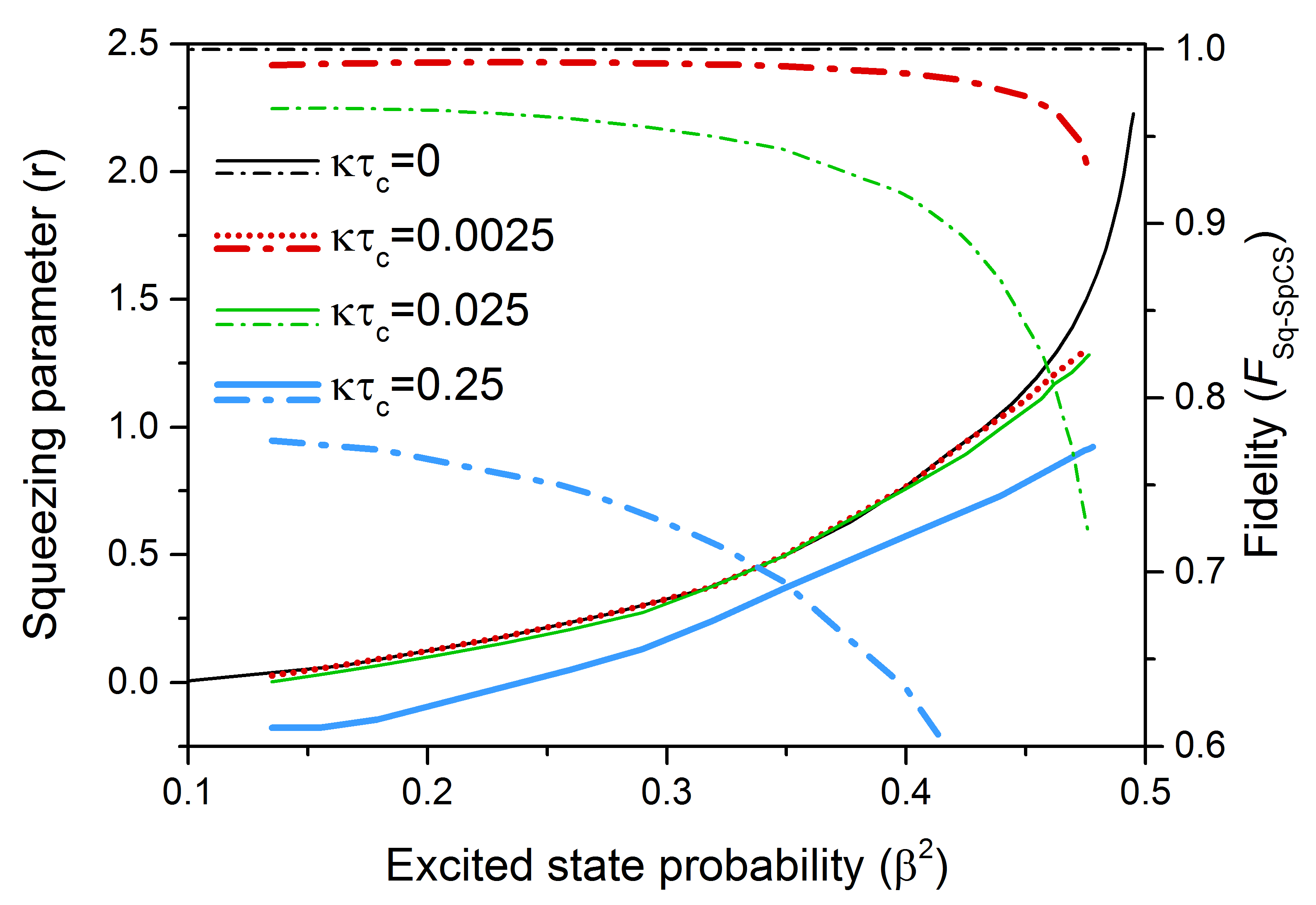}
\caption{(Color online) Squeezing parameter $r$ (solid lines or dotted lines) and fidelity $F_{\rm Sq-SpCS}$ (dash-dot lines) with respect to a Sq-SpCS as a function of excited state probability $\beta^2$ of atoms.
As denoted in the legend, different colors (grayscales) and line styles represent different cavity decay rate $\kappa$ times the characteristic time $\tau_c$. 
The fidelity $F_{\rm Sq-SpCS}$ is always higher than $F_{\rm SpCS}$ in Fig.\ \ref{fig2} for the same condition. The main obstacle for the high fidelity is the cavity decay rate.}
\label{fig3}
\end{figure}

If we compare the SPSb-SqVS state with a Sq-SpCS, we obtain the squeezing parameter and its fidelity $F_{\rm Sq-SpCS}$ with respect to the Sq-SpCS as shown in Fig.\ \ref{fig3}. The Sq-SpCS used for comparison is the one that maximizes the fidelity with the given  SPSb-SqVS. 
Particularly, for infinitesimally small cavity decay, $F_{\rm Sq-SpCS}$ is very close to unity.
If the excited state probability is larger than 0.5, there is no steady state because the emission of excited state atoms is always stronger than the absorption of ground state atoms regardless of average photon number of the cavity field.

\section{Effects of substantial cavity decay} \label{sec4}

In order to consider a substantial cavity decay rate and dynamics of the system, we adopt QTS analysis (Secs.~\ref{sec4} and \ref{sec5}). 
In QTS, time evolution of the total atom-field wave function is calculated under the Tavis-Cummings Hamiltonian while the wave function is made to collapse upon cavity decay and exit of atoms.
Additional assumptions for the simulation are as follows.
\begin{itemize}
\item Atomic interaction with environment, {\it i.e.} atomic free-space spontaneous emission, is negligible compared to both atom-cavity interaction and cavity decay. 
\item Atom injection time for each atom is random.
\item The superposition-state phase of a newly prepared atom is opposite to that of the atom injected just before.
\item The velocity as well as the atom-field coupling are the same for all atoms.
\item The cavity field profile along the atomic motion is a top-hat shape.
\end{itemize}

\begin{figure}
\includegraphics[width=3.4in]{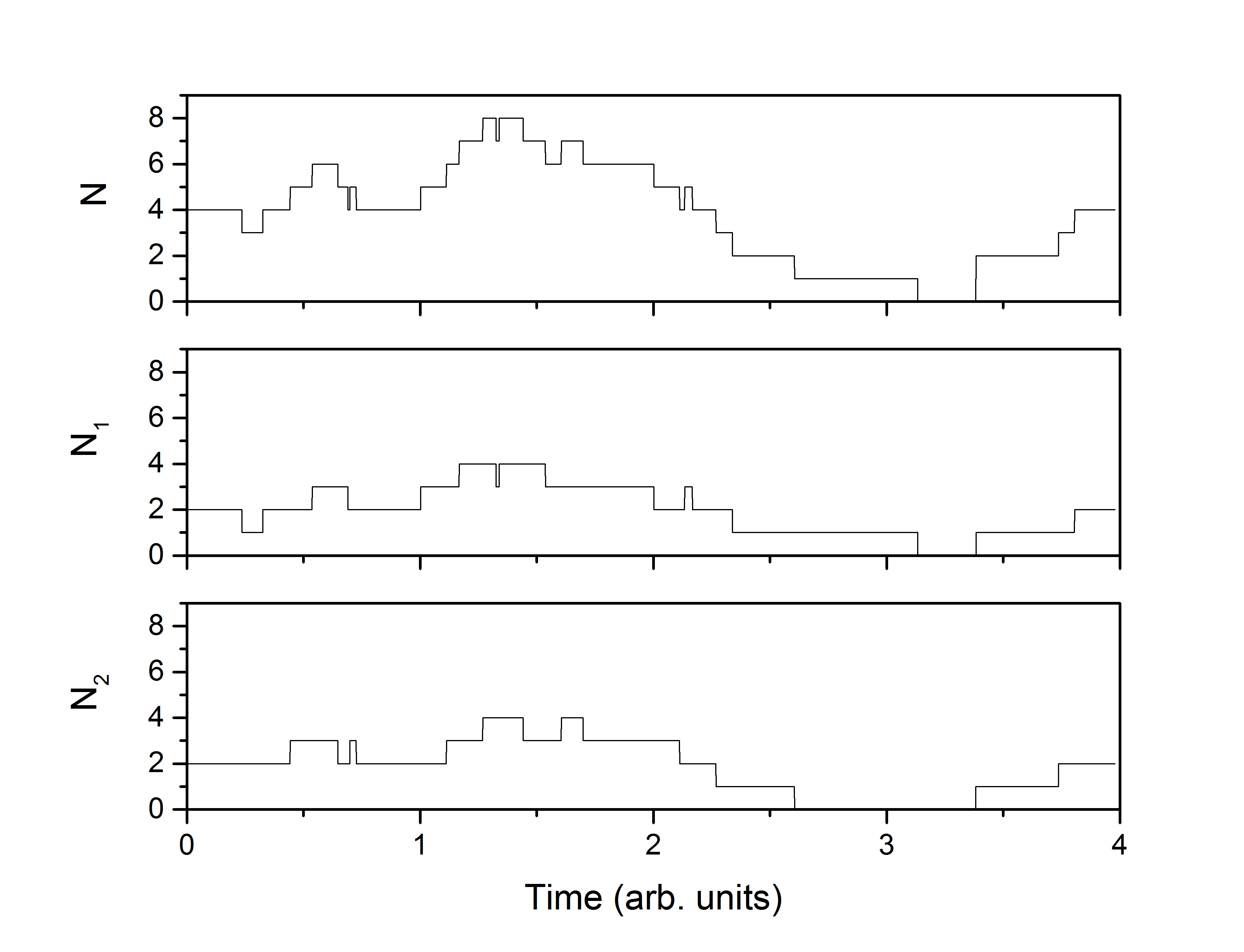}
\caption{Number of atoms in the cavity in our QTS calculations. The symbol
$N$ is the total number of atoms in the cavity and $N_1$($N_2$) represents 0($\pi$) phase atoms in the cavity. Time is rescaled to $t_{\rm int}$. Because atomic injection time for each atom is random, the numbers of atoms $N, N_1$ and $N_2$ fluctuate around their mean values. The time sequences are taken from the QTS results corresponding to $\langle N\rangle=2$, $\beta^2=0.41$ and thus $\kappa\tau_c= 0.0025$ in Fig.\ \ref{fig2}.}
\label{fig4}
\end{figure}

In contrast to the analytic calculation in the preceding sections, here we inject each atom into the cavity randomly in time in order to simulate experimental situations. 
As a result, both the numbers $N_1$ and $N_2$ of atoms in the cavity with opposite phases fluctuate in time as shown in Fig.\ \ref{fig4}, which depicts one of the time sequences of QTS results. 
Analytic calculations considering the non-simultaneous atomic injection are presented in Appendix \ref{appen:B}.

The formation of a SqVS in the cavity is found to be robust against 
imperfect detection efficiency of the photons decaying out of the cavity: missed detection events act as weak perturbations, not affecting the rapid formation of a SqVS in the cavity. 
For simplicity in the analysis below, we can thus assume that the overall photon-detection efficiency is nearly zero. 
The results would be worse than, but not much different from an ideal case of 100\% photon-detection efficiency. 
In practice, imperfect photon detection happens because of mirror scattering and absorption losses, low detector quantum efficiency 
and loss along the optical path.
Under substantial cavity decay, the SqVS would undergo significant decoherence and the SPSb-SqVS shows a lower photon number and a worse fidelity than the ideal case of the infinitesimally small cavity decay.

In the limit of $gt_{\rm int} \ll 1$, the rate of SqVS formation can be expressed as the inverse of a characteristic time $\tau_c$, which is given by (see Appendices {\ref{appen:A} and {\ref{appen:B})
\begin{equation}
1/\tau_c = e^{-2r}\langle N \rangle g^2t_{\rm int}
\end{equation}
where $\langle N\rangle$ is the average number of atoms in the cavity and it can be much larger than unity. 
It should be noted that the expression for $1/\tau_c$ is like the laser gain except for the squeezing factor $e^{-2r}$. 
The degree of SqVS decoherence can then be expressed as a ratio between cavity decay rate $\kappa$ (loss) and the SqVS formation rate $1/\tau_c$ (gain), or $\kappa \tau_c$.

Average photon numbers decreased due to substantial cavity decay are shown in Fig.\ \ref{fig2} as a function of $\kappa \tau_c$. 
Likewise, 
the fidelity with respect to a SpCS (Fig.\ \ref{fig2}) as well as the maximal fidelity with respect to the Sq-SpCS (Fig.\ \ref{fig3}) decrease as $\kappa \tau_c$ increases due to the decoherence in the formation of the SqVS.
As mentioned above, a SPSb-SqVS is similar to a Sq-SpCS.
Therefore, the fidelity with respect to a SpCS decreases much more than that to a Sq-SpCS as the pumping parameter increases in the case of substantial cavity decay.
Nevertheless, a mean photon number larger than unity as well as the fidelity $F_{\rm Sq-SpCS}$ larger than 0.9 can be obtained with a substantial cavity decay rate of $\kappa \tau_c=0.025$, where $\tau_c$ can be made fairly small by increasing the average number $\langle N \rangle$ of atoms.

\section{Dynamics} \label{sec5}

After a cat-like state is obtained by a photon click, the cavity field will be restored to the SqVS by the atom-field interaction in time.
Restoration rate for the SqVS is proportional to $1/\tau_c$ in the limit of $gt_{\rm int} \ll 1$.
In order to see the restoration process, we calculated the time evolution of $F_{\rm SqVS}$ after a photon click occurs.
Time evolution of the fidelities and the average photon number for representative $\beta^2$ values are plotted in Fig.\ \ref{fig5}(a), showing that a SqVS is indeed recovered in a characteristic time $\tau_c$.

\begin{figure}
\includegraphics[width=3.4in]{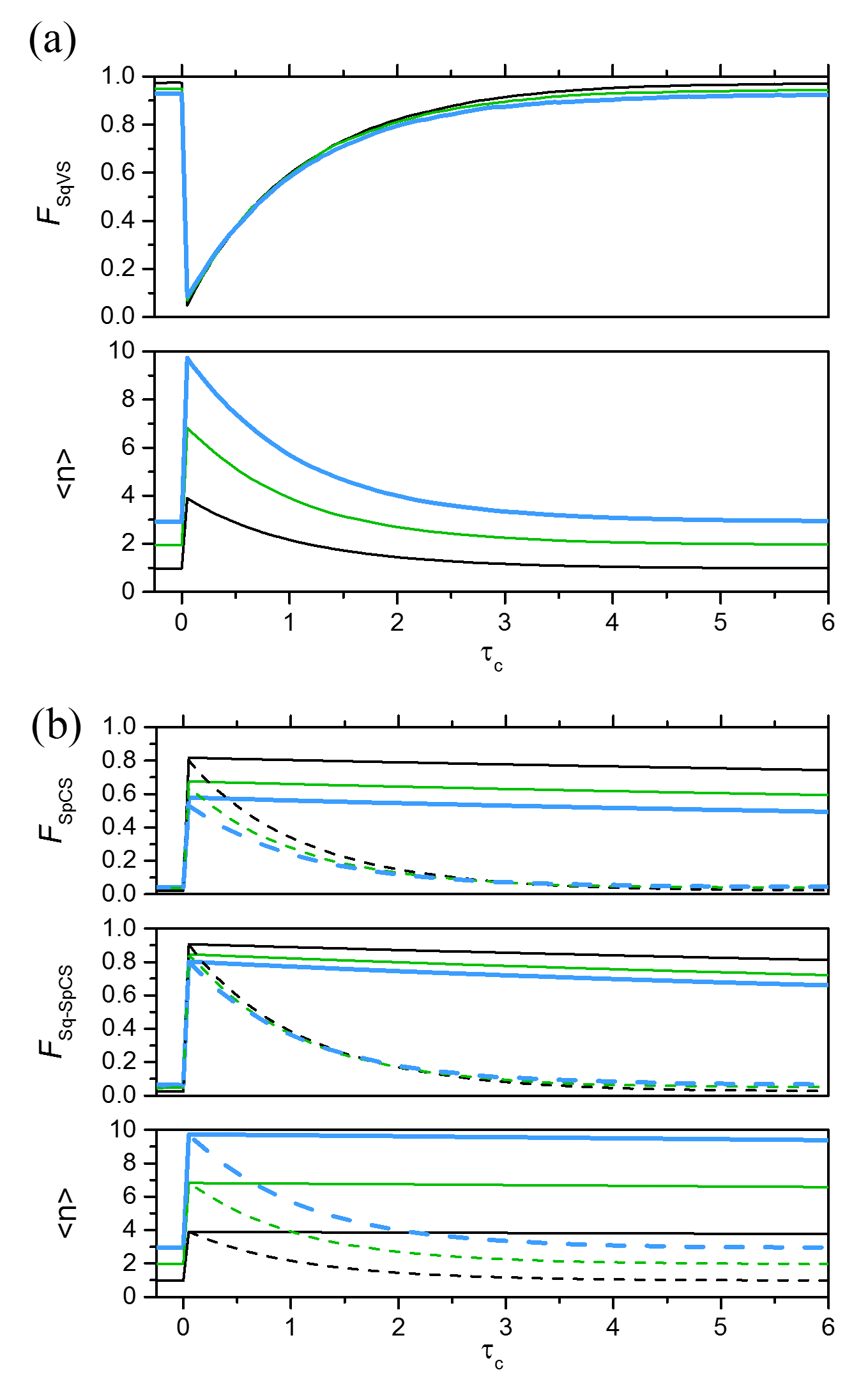}
\caption{(Color online) Time evolution and decoherence of the cavity field with various excited state probability $\beta^2$ of atoms. 
We assume $gt_{\rm int}=0.1$ and $\kappa\tau_c=0.025$.
The values of $\beta^2$ are set to generate the following average photon number of SqVSs: $\langle n_{\rm SqVS} \rangle$ = 1.0 (black line), 2.0 [green (thin gray) line] and 3.0 [blue (thick gray) line].
(a) The fidelity $F_{\rm SqVS}$ is obtained with respect to the SqVS having the same average photon number as the state before a photon click occurs.
When $t=0$, a photon click occurs and the SqVS jumps to a cat-like state.
Regardless of $\langle n_{\rm SqVS} \rangle$, the SqVSs is restored within the time $\tau_c$.
(b) Decoherence of the cavity field with (dashed lines) and without (solid lines) the atom-field interaction.
The fidelity $F_{\rm SpCS}$ is obtained with respect to the usual SpCS having the same average photon number as the state right after a photon click. 
The fidelity $F_{\rm Sq-SpCS}$ is calculated with respect to the one that maximizes the fidelity for a given SPSb-SqVS.
Because $1/\tau_c$ is much larger than $\kappa$ in our example, the decoherence rate of the cat-like state in the presence of the atom-field interaction is much faster than the case without the atom-field interaction.
}
\label{fig5}
\end{figure}

However, the restoration of the SqVS by the atom-field interaction can be an obstacle for observing a cat-like state for a long time. In general, the cat-like state can be destroyed by both cavity decay and atom-field interaction. The decoherence rate of the cat-like state by the atom-field interaction is proportional to $1/\tau_c$, so is the restoration rate of the SqVS. The decoherence rate by the cavity decay is proportional to $\kappa$, like a usual Schr\"odinger cat in a cavity. Figure \ref{fig5}(b) shows the decoherence of the cat-like state for two cases - one with and the other without the atom-field interaction.
Since our scheme is limited to the case where the squeezed state formation rate (the gain) for the field is much larger than the cavity decay rate (the loss) just like a laser, the atom-field interaction is the main source of decoherence for the cat-like state. This decoherence, however, can be eliminated by disabling the atom-field interaction, for example, by stopping atom injection, transferring the injected atoms to a dark state or Stark-shifting the atomic transition line.

\begin{table*}\label{table1}
\begin{tabular}{c | c | c | c | c | c | c | c}
\setlength{\tabcolsep}{40pt}
Parameter & set 1 & set 2 & set 3 & set 4 & set 5 & set 6 & remark \\ 
\hline\hline
Mirror finesse (million) & 1.0 & 2.0 & 4.0 & 8.0 & 16.0 & 64.0 &\\
$\langle N \rangle$ & 8 & 8 & 8 & 8 & 8 & 8 & \\
Cavity length (mm) & 5.0 & 5.0 & 5.0 & 5.0 & 5.0 & 5.0 & assumed \\ 
Velocity of atoms & 400 & 400 & 400 & 400 & 400 & 400 &parameters\\
Excited state Probability	&	0.34	&	0.37	& 0.39 	&	0.40	&	0.41	& 0.47 &\\ 
\hline
$gt_{\rm int}$	&	0.25	&	0.25	&	0.25	&	0.25	& 0.25 & 0.25 & derived \\ 
$\kappa\tau_c$	&	0.15	&	0.096	& 0.056	&	0.033	&	0.018 &	0.012 & parameters \\ 
\hline
Expected size of cat ($\langle n \rangle$) & 1.73 & 2.31 & 2.83 & 3.47  & 3.86 & 10.79 & \\ 
Fidelity ($F_{\rm SpCS}$)  & 0.84 & 0.84 & 0.85 & 0.84  & 0.84 & 0.59 & $\eta$=0.5 \\
Fidelity ($F_{\rm Sq-SpCS}$) & 0.85 & 0.87 & 0.89 & 0.91  & 0.93 & 0.86 & \\
\hline
Expected size of cat ($\langle n \rangle$) & 1.72 & 2.25 & 2.74 & 3.38  & 3.79  & 10.27 & \\ 
Fidelity ($F_{\rm SpCS}$)  & 0.93 & 0.92 & 0.91 & 0.89  & 0.87 & 0.64 & $\eta$=1.0 \\
Fidelity ($F_{\rm Sq-SpCS}$) & 0.94 & 0.95 & 0.96 & 0.96  & 0.96 & 0.92 & \\
\hline
\end{tabular}
\caption{Suggested experimental parameters for $^{1}$S$_{0}$-$^{3}$P$_{1}$ transition of $^{174}$Yb at $\lambda=555.6$nm with an excited-state decay rate of 183kHz. The radius of mirror curvature is assumed to be 10mm. Overall photon-detection efficiency 
is denoted as $\eta$. Finesse of mirrors in an optical regime can reach a few million with the current supermirror technology \cite{supermirror}. Narrow velocity distribution of atoms can be realized by resonantly deflecting an atomic beam with a laser \cite{atomic beam}. Moreover, it is not technically difficult to achieve the listed or even larger $\langle N \rangle$. Excited state probability can be controlled by an optical pumping. 
}
\end{table*}

\section{Suggested experiments} \label{sec6}
By using QTS, we have confirmed that cat-like states are formed under the experimental conditions which are within the reach of the present experimental capacity. 
Table 1 summarizes some numerical simulation results done for $^{174}$Yb. 
Even though a larger intracavity atom number $\langle N \rangle$ is expected to increase the size and the fidelity of the cat-like state, exponential increase in computation time, proportional to $\sim 2^{\langle N \rangle}$, limits the largest $\langle N \rangle$ that can be explored in Table 1. 
Instead of increasing $\langle N \rangle$, we increased the mirror finesse $\mathcal{F}$ in the simulation in order to obtain large cat-like states.
This practice is allowed since the formation of a SqVS is mainly determined by $\kappa \tau_c$, which is inversely proportional to $\mathcal{F}  \langle N\rangle$.
For example, parameter set 5 is equivalent to a set with $\mathcal{F}=10^6$ and $\langle N\rangle=128$ with the other parameters the same. These parameters are experimentally achievable with the present technology and the resulting cat-like state would have high fidelity and a large size.
Figure \ref{fig6} graphically shows the average photon number and the fidelity of the cat-like state under various excited state probability and $\mathcal{F} \langle N \rangle$ values.
We can see that cat-like states with the average photon number $\langle n \rangle$ larger than 10 can be achieved under certain conditions.
In the previous free-propagating cat-like-state experiments, on the other hand, the size of cat-like states made by single-photon subtraction was limited to $\langle n \rangle \sim 2$  \cite{large_cat} 
due to technical reasons associated with
nonlinear crystals \cite{pump_induced_absorption, squeezed_state_recent_review}.
Our method is free from such nonlinear crystal issues and is thus expected to produce larger cat-like states.

\begin{figure}
\includegraphics[width=3.4in]{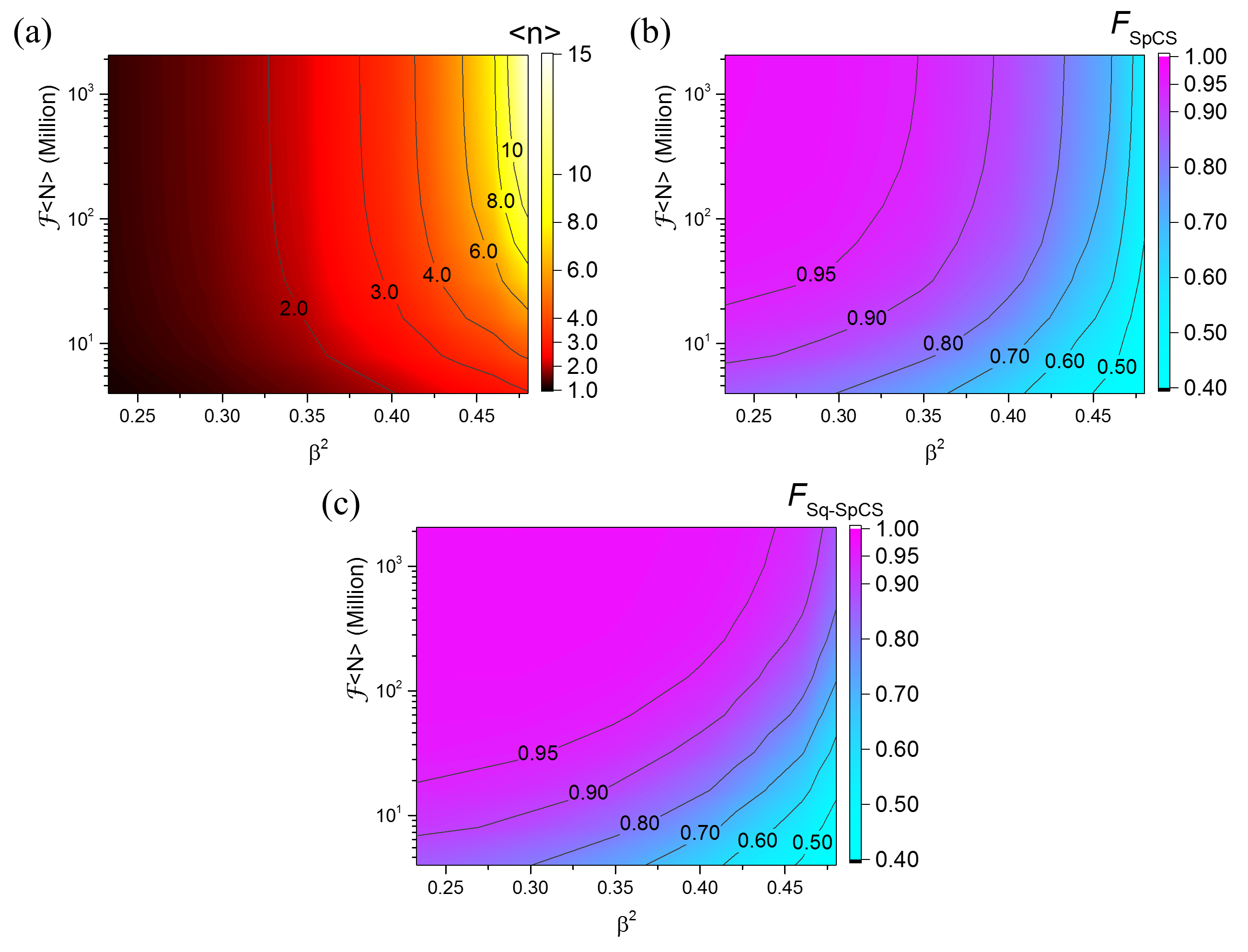}
\caption{(Color online)
(a) Average photon number $\langle n\rangle$,
(b) fidelity $F_{\rm SpCS}$  and
(c) fidelity $F_{\rm Sq-SpCS}$ of the cat-like state as a function of excited state probability $\beta^2$ and $\mathcal{F}\langle N \rangle$. 
The tendency is similar to that in Fig.\ \ref{fig2}.  
The other parameters are the same as the set 1 in Table \ref{table1}.
Overall photon-detection efficiency is assumed to be 0.5.}
\label{fig6}
\end{figure}  

\begin{figure}
\includegraphics[width=3.4in]{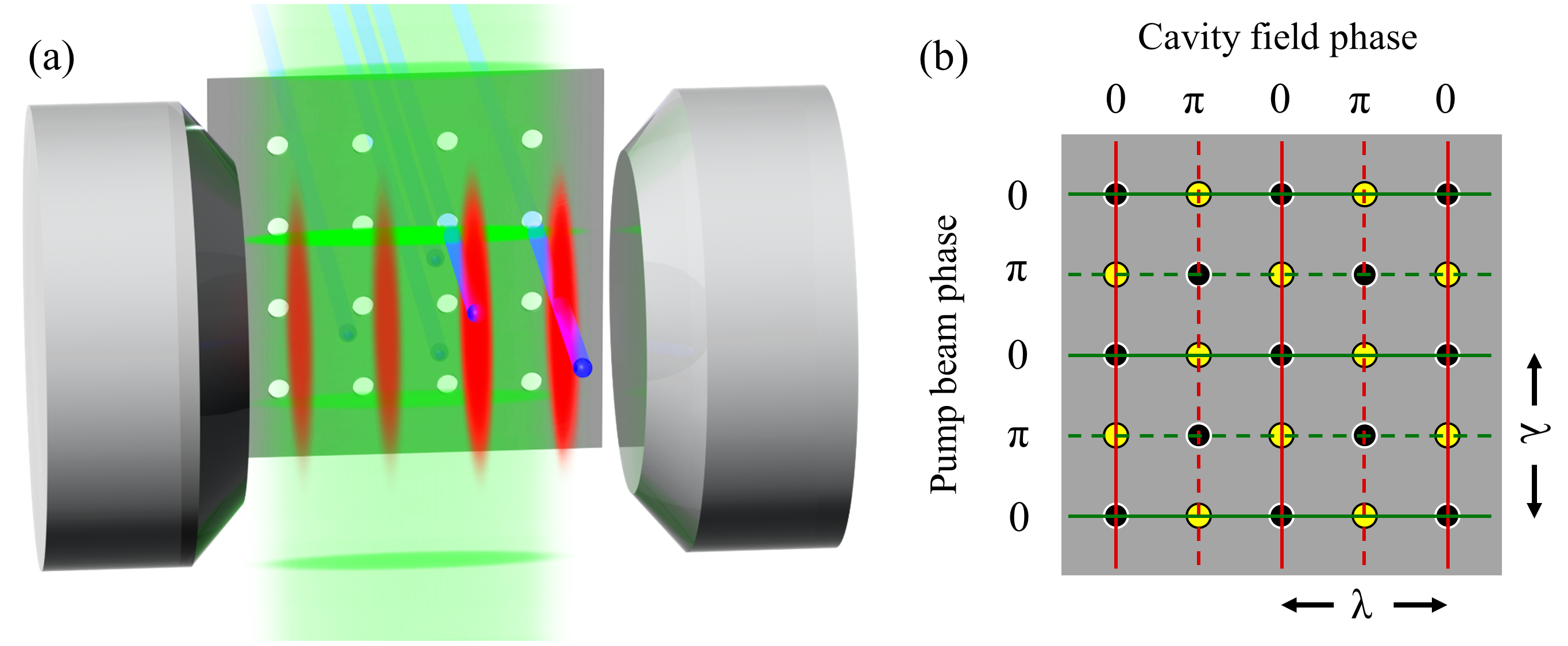}
\caption{ (Color online)
Experimental scheme for injecting opposite phase atomic dipoles using a nanohole array.
(a) Randomly distributed atoms (blue spheres) are filtered by the nanohole square array with a $\lambda/2$ period.
Equiphase planes of the pump beam (translucent green) are denoted in solid green. Antinodes of the cavity field are denoted in red.
(b) A square-lattice nanohole array can prepare atoms in two groups: one group with 0 phase (e.g., black holes) and the other group with $\pi$ phase (yellow holes). 
Solid green horizontal (red vertical) lines indicate the equiphase planes of the pump beam (cavity field) with phase 0 while the dashed green horizontal (red vertical) lines indicate the equiphase planes of the pump beam (cavity field) with phase $\pi$.
Atoms separated by $\lambda/2$ in the pump beam direction experience a phase difference $\pi$ of the pump field while the atoms separated by $\lambda/2$ in the cavity axis direction experience an identical amplitude but opposite phases in the atom-field coupling.
With these two effects combined, atoms can effectively interact with the cavity field as in the superposition states with opposite phases. }
\label{fig7}
\end{figure}  

Injecting atomic dipoles with opposite phases can be achieved by using a nanohole array \cite{nanohole-14} with a $\lambda/2$ period as an atomic beam aperture (Fig.\ \ref{fig7}). 
Such a nanohole array can localize the atomic position with a $\lambda/2$ period, tuned to the cavity antinodes.
Atoms separated by a distance $\lambda/2$ in the direction of a pump laser beam experience a phase difference of $\pi$ for the transverse pump laser and are prepared in opposite-phase atomic dipoles. 
As $\langle N \rangle$ increases, the relative number difference between opposite phase atomic dipoles would decrease as $\sim 1/\sqrt{\langle N \rangle}$ approaching an ideal situation discussed in Secs. V and VI.

Measurement of the nonclassical cavity field state can be achieved by using dispersive atom-field interaction \cite{Brune-92, Del-08}, usual homodyne measurement \cite{homodyne-1, homodyne-2} or unbalanced homodyne measurement \cite{unbalanced_homodyne}. In particular, the photon-counting-based unbalanced homodyne technique would allow detection of both a herald event and the consequent homodyne counting measurements in a simple detector configuration.

\section{Conclusion}\label{sec7}

We have proposed and analyzed a novel way of generating an optical Schr{\"o}dinger's cat-like state in a cavity by injecting opposite phase atomic dipoles even in the presence of a substantial cavity decay. 
Despite the significant cavity decay, formation of Schr{\"o}dinger's cat-like state is confirmed with QTS if the condition $\kappa\tau_c \ll 1$ is satisfied.
Under this condition, the cavity field converges to a SqVS in a characteristic time $\tau_c$. 
The squeezed vacuum state then collapses to a Schr{\"o}dinger's cat-like state by a single cavity photon decay, which acts as single-photon subtraction operation on the SqVS.
After the cat state is generated, atom-field interaction restores the same SqVS as formed before in a characteristic time $\tau_c$ while the Schr{\"o}dinger's cat-like formation is heralded again by observing a single photon click outside the cavity.
We propose possible experiments within the reach of the current technology based on our QTS analysis on ytterbium atoms.
Our approach is distinct from the previously known methods in that a Schr{\"o}dinger-cat-like state is rapidly formed in a laser-like setting without complicate operations in a cavity and in that its size is relatively large in the optical region.
Such a cat-like state is known to be useful in quantum information processing \cite{QI-1,QI-2,QI-3} and in quantum metrology \cite{Quantum metrology-1,Quantum metrology-2,Quantum metrology-3}.

\acknowledgements
We thank H.\ Jeong for helpful comments. This work was supported by a grant from Samsung Science and Technology Foundation under Project Number SSTF-BA1502-05.

\appendix

\section{Derivation of steady-state formation of the squeezed vacuum state}
\label{appen:A}
In order to show the squeezed vacuum state formation, let us consider a time evolution operator of Tavis-Cummings Hamiltonian without cavity decay terms. The time evolution operator is given by,
\begin{equation}
U(t)=\exp\left[-igt\left( aJ_++a^\dagger J_-\right)\right]
\end{equation}
where $J_+=\sigma_1^\dagger+\sigma_2^\dagger$, $J_-=\sigma_1+\sigma_2$ and $\sigma_i = |\downarrow\rangle\langle\uparrow|$ is lowering operator of $i$ th atom. We can generalize the time evolution operator for $N$ pairs of atoms present at the same time as,
\begin{equation}
U(t)=\exp\left[-igt\sum_i^N (aJ_{i+}+a^\dagger J_{i-})\right]
\end{equation}
When $gt_{int} \ll 1$ is satisfied, the state vector is transformed after the specific time $t_{\rm int}$ to
\begin{eqnarray}
|\psi(t_{\rm int})\rangle &=& \left\{ 1-igt_{\rm int}\sum_i^N (aJ_{i+}+a^\dagger J_{i-}) \right. \nonumber\\
& & - \left. \frac{1}{2}g^2t_{\rm int}^2\left[\sum_i^N (aJ_{i+}+a^\dagger J_{i-})\right]^2 \right. \nonumber\\
& &+ \left. \mathcal{O} \left( g^{3}t_{\rm int}^{3}\right)\right\} \psi(0)\rangle
\end{eqnarray}
Tracing over atomic states would result in a density matrix for the field as
\begin{equation}
\rho_{\rm field}(t_{\rm int})={\rm Tr}_{\rm atom}\left[|\psi(t_{\rm int})\rangle \langle\psi(t_{\rm int})|\right]
\end{equation}
Some calculations yields
\begin{eqnarray}
& &\left[\rho_{\rm field}(t_{\rm int})-\rho_{\rm field}(0)\right]/t_{\rm int} \nonumber\\
& &=2 g^2 t_{\rm int}N (\alpha^ 2 a - \beta^2 a^\dagger )\rho_{\rm field} (0) (\alpha^ 2 a - \beta^2 a^\dagger )^\dagger \nonumber \\
& &\;\; \; - g^2 t_{\rm int}N (\alpha^2 a^\dagger - \beta^2 a )(\alpha^ 2 a - \beta^2 a^\dagger) \rho_{\rm field} (0) \nonumber\\
& & \;\;\; - g^2 t_{\rm int}N \rho_{\rm field} (0)\left[ (\alpha^2 a^\dagger - \beta^2 a ) (\alpha^ 2 a  - \beta^2 a^\dagger ) \right] ^\dagger \nonumber\\
& &\;\;\;+ \mathcal{O} \left( g^{4}t_{\rm int}^{3}\right) 
\end{eqnarray}
Because we are assuming the case of $gt_{\rm int} \ll 1$, we can replace $[\rho_{\rm field}(t_{\rm int})-\rho_{\rm field}(0)]/t_{\rm int}$ with $\dot {\rho}_{\rm field}(t_{\rm int})$. In order to cast the result in a more intuitive way, let us express the time evolution of the state in a squeezed vacuum state basis, $\rho_{\rm field} (t) = \hat{S}(\xi)\rho'(t)\hat{S}^\dagger(\xi)$. Here, the squeezing operator is defined as $\hat{S}(\xi)\equiv\exp\left(\frac{1}{2}\xi^*a^2-\frac{1}{2}\xi a^{\dagger}{}^2\right)$ with a squeezing parameter $r=| \xi |=\tanh^{-1} (\beta^2/\alpha^2)$.
By using the following properties,
\begin{align}
&(\alpha^ 2 a - \beta^2 a^\dagger )\hat{S}(\xi) =  \hat{S}(\xi) a \nonumber \\
&(\alpha^2 a^\dagger - \beta^2 a )\hat{S}(\xi) = \hat{S}(\xi) a^\dagger  \nonumber
\end{align}
the equation can be rewritten as,
\begin{eqnarray}
&\hat{S}&(\xi)\dot{\rho}'(t)\hat{S}^\dagger(\xi)\nonumber\\
&=&\hat{S}(\xi)\frac{1}{2\tau_c}\left(2a\rho'(t)a^\dagger-a^\dagger a \rho'(t)-\rho'(t)a^\dagger a \right)\nonumber\\
& & \hat{S}^\dagger(\xi)
\end{eqnarray}
where, $1/\tau_c\equiv 2e^{{-2|\xi|}}Ng^2t_{\rm int}$. By dividing both side $\hat{S}(\xi)$ and $\hat{S}^\dagger(\xi)$, the equation is the cavity field decay equation at zero temperature with a decay rate $1/\tau_c$. So the density matrix evolves to,
\begin{equation}
{\rho}_{\rm field}(t)=\hat{S}(\xi)\rho'(t)\hat{S}^\dagger(\xi) \rightarrow \hat{S}(\xi)\left|0 \rangle \langle 0\right|\hat{S}^\dagger(\xi)
\end{equation}
As a result, the density matrix goes to the SqVS in a characteristic time $\tau_c$.

\section{Generalization to non-simultaneous injection of atoms}
\label{appen:B}
In the main text and in Appendix \ref{appen:A}, we assume that two opposite-phase atoms are simultaneously injected. For a non-simultaneous injection case, let us now consider a time difference $\Delta t$ between two opposite phase atoms. The time evolution operator of the non-simultaneous injection case can be expressed as,
\begin{eqnarray}
U(t)&=&\exp\left[-ig\Delta t(a\sigma_2^{\dagger}+a^{\dagger}\sigma_2)\right]\nonumber\\
& &\exp\left[-ig(t_{\rm int}-\Delta t)(a(\sigma_2^{\dagger}+\sigma_1^{\dagger})+a^{\dagger}(\sigma_2+\sigma_1))\right] \nonumber \\
& &\times\exp\left[-ig\Delta t(a\sigma_1^{\dagger}+a^{\dagger}\sigma_1)\right]
\end{eqnarray}
where $\sigma_1$($\sigma_2$) is a lowering operator of first(second) injected atom and $\Delta t$ is less than or equal to $t_{\rm int}$. As in appendix \ref{appen:A}, the state vector is transformed after the specific time $t_{\rm int}$ to
\begin{eqnarray}
& &|\psi(t_{\rm int}+\Delta t)\rangle\nonumber\\
& &\;\;= \Bigg\{ 1-igt_{\rm int} (aJ_{i+}+a^\dagger J_{i-}) \nonumber\\
& &\;\;- \frac{1}{2}g^2t_{\rm int}^2\left( aJ_{i+}+a^\dagger J_{i-}\right)^2  \nonumber \\
& &\;\;-\frac{1}{2}g^2\left( 2t_{\rm int}{\Delta t}-{\Delta t}^2 \right) \left( \sigma_2^\dagger \sigma_1 - \sigma_2 \sigma_1^\dagger \right) \nonumber\\
& &\;\;+ \mathcal{O} \left( g^{3}t_{\rm int}^{3}\right)\Bigg\} | \psi(0)\rangle
\end{eqnarray}
Tracing over atomic states yield the same result as in Appendix \ref{appen:A} because the fourth term in the curly brackets does not have field annihilation or creation operator. So we can generalize the result as,
\begin{equation}
1/\tau_c \approx 2e^{-2|\xi|}Ng^2t_{\rm int}=e^{-2|\xi|}\langle N \rangle g^2t_{\rm int}
\end{equation}
where $N$ is the number of pairs of opposite-phase atoms and $\langle N \rangle$ is the average number of atoms in the cavity.

\end{document}